\documentclass[10pt,letterpaper,twocolumn]{article} %% two column, final layout
\usepackage{epsfig}
\usepackage{graphicx}
\usepackage{amssymb}
\usepackage{amsmath}
\usepackage{ol2}

\usepackage{color}
\newcommand{\ha}{\hat{a}}
\newcommand{\hb}{\hat{b}}

\begin{document}
\twocolumn[
\title{Mode-resolved Photon Counting via Cascaded Quantum Frequency Conversion}

\author{Yu-Ping Huang and Prem~Kumar}

\address{Center for Photonic Communication and Computing, EECS
Department, Northwestern University, Evanston, IL 60208-3118}

%\email{kumarp@northwestern.edu}
\begin{abstract}
Resources for the manipulation and measurements of high-dimensional photonic signals are crucial for implementing qu$d$it-based applications. Here we propose potentially high-performance, chip-compatible devices for such purposes by exploiting quantum-frequency conversion in nonlinear optical media. Specifically, by using sum-frequency generation in a $\chi^{(2)}$ waveguide we show how mode-resolved photon counting can be accomplished for telecom-band photonic signals subtending multiple temporal modes. Our method is generally applicable to any nonlinear medium with arbitrary dispersion property.
\end{abstract}
]
\maketitle
Photonic signals are widely employed in many modern applications such as secure key generation, quantum computation, and near Holevo-capacity telecommunications \cite{Rev-Qua-cryp}. In these applications, photons occupying low-dimensional Hilbert spaces composed of polarization or spatial degrees of freedom are most commonly used. Recently, the use of high-dimensional photonic signals, known as qu\textit{d}its, has attracted attention as they can offer significant advantages over qubits in terms of higher channel capacity, better information security, and so on. There have been quite a few proposals for using photonic qu{\it d}its in applications such as secure telecommunications \cite{PhysRevLett.89.197901} and super-dense coding \cite{PhysRevA.66.014301}.

Crucial to implementing the aforementioned qu$d$it-based applications are the capabilities for manipulating and measuring high-dimensional photonic signals. Existing solutions have been based on serializing multiple, linear, two-port optical devices \cite{KwiatHyperEntangled05} or using lossy spatial-mode modulators \cite{OAMQudits11}. Their performance in practice, particularly in large-\textit{d} systems, will be highly inefficient due to the need for complicated experimental setups with high concomitant transmission losses.

In this Letter, we study a potentially high efficiency, low-loss approach for manipulating and measuring photonic signals in high-dimensional Hilbert spaces composed of polarization, spatial, and temporal degrees of freedom. It is built on quantum-frequency conversion (QFC), a coherent process that can translate the carrier frequency (wavelength) of photonic signals without disturbing their quantum states, including any correlation or entanglement with other quantum objects \cite{Kumar90}. It is well known that QFC in nonlinear media is efficient only for those appropriate combinations of wavelengths, spatiotemporal modes, and polarizations of the interacting light waves for which phase matching (PM) occurs. Thus it is possible to engineer the conditions for PM such that only photons in desired modes are frequency converted with high efficiency, while those in other modes remain unaffected \cite{QPG_SilberhornOE11}. By detecting the frequency-converted signals with ordinary single-photon detectors, differences in the photons' mode-profile details, which are otherwise difficult to monitor, can be precisely measured. In addition, by using photon-number-resolving (PNR) detectors, mode-resolved photon counting (MRPC) can be accomplished, based on which quantum statistics of high-dimensional photonic signals can be characterized on a mode-by-mode basis. Because the PM conditions can be engineered by tailoring the pump waves driving the QFC process, ultrafast manipulations and measurements of qu\textit{d}its are realizable in chip-scale devices with low transmission losses \cite{QFCFWMPRA12}. Our approach thus holds promise for use in practical applications of high-dimensional photonic signals.

For concreteness, we focus specifically on the MRPC application. Keeping telecom-band signals in mind, we explicitly consider sum-frequency generation (SFG) that converts near-infrared photons into the visible, where the converted photons can be detected with high efficiency using silicon-based avalanche photodiodes (APDs) or PNR detectors \cite{QFC_MaOE09}. In our design, multiple QFC processes are applied sequentially to the input multimode signal, each providing efficient conversion to a single mode that is orthogonal to the rest. In each QFC stage, the converted sum-frequency (SF) light is detected using an APD or a PNR detector. By modulating the pump pulses, mutually-unbiased conversion-mode bases can be realized, allowing implementation of quantum protocols with multilevel encodings \cite{BBN1}.

\begin{figure}
   \centering \epsfig{figure=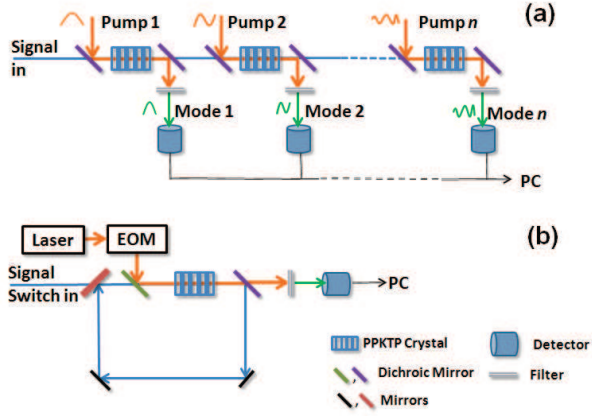,width=8.0 cm}
   \caption{Schematic of MRPC in a serial (a) and an optical-loop (b) realization.}  \label{fig1}
\end{figure}

A schematic of our proposed system is shown in Fig.~\ref{fig1}(a), where the optical signals to be measured pass through a series of identical SFG waveguides, each pumped by pulses of appropriate temporal shape to drive one of the orthogonal conversion modes. The converted SF light after each waveguide is separated using a dichroic mirror and detected with an APD or a PNR detector. This particular setup utilizes multiple crystals and detectors. A more economical system is possible that uses only one crystal and one single-photon detector by adopting the optical-loop configuration shown in Fig.~\ref{fig1}(b). We note that since the conversion modes, and hence the detection modes, are defined by the pump pulses, very short detection windows (down to tens or hundreds of femtoseconds) can be realized with ultralow timing jitter (tens of femtoseconds or less).

In the following, we focus on MRPC in the temporal domain, assuming a single spatial mode for each light wave. Extending the present analyses to include multiple spatial modes is straightforward \cite{QFC_Image12} and will be presented elsewhere. For concreteness, we explicitly consider the following coupled Heisenberg equations of motion that describe the propagation and interaction of light waves in a lossless $\chi^{(2)}$ waveguide:
\begin{subequations}
\label{eq1}
\begin{align}
& &(\partial_z+\mu\partial_t)\ha(z,t)= i\eta f(t) \hb(z,t),~~  \\
& &(\partial_z+\nu\partial_t)\hb(z,t)= i\eta^\ast f^\ast(t)\ha(z,t).
   \end{align}
\end{subequations}

In Eqs.~(\ref{eq1}), which are derived in a moving frame traveling with the pump wave, $\hat{a}$ and $\hat{b}$ are the annihilation operators for the signal and the SF waves, respectively, satisfying the standard commutation relations, viz., $[\hat{a}(z,t),\hat{a}^\dag(z',t')]=\delta(z-z')\delta(t-t')$; $\eta$ is a coefficient measuring the strength of the SFG interaction including the effect of spatial modal overlap; $f(t)$ gives the temporal profile of the pump pulses; and $\mu$ and $\nu$ are ``slownesses'' (inverse group velocities) of the signal and SF waves, respectively, relative to the pump wave.

Equations~(\ref{eq1}) can be solved following the standard Green's function approach. When the input SF-wave is in the vacuum state, the output at the SF wavelength can be obtained through the following transformation
\begin{equation}
\label{eq2}
    \hat{b}(t)=\int d t' G(t,t') \hat{a}(t')+\hat{\xi}(t),
\end{equation}
where $G(t,t')$ is a Green's function determined by Eqs.~(\ref{eq1}), and $\hat{\xi}(t)$ describes potential in-coupling of quantum noise due to, for example, transmission loss and spontaneous Raman scattering \cite{QFC_MaOE09}. Applying normal-mode decomposition similarly to the treatment of free-space diffraction (see \cite{Shapiro09} and references therein), $G(t,t')$ can be rewritten as
%\begin{equation}
%\label{eq3}
    $G(t,t')=\sum^{\infty}_{n=0} \lambda_n \phi_n(t)\psi_n^\ast(t')$,
%\end{equation}
where $\psi_n(t)$ and $\phi_n(t)$ are the $n$-th pair of input-output modes satisfying $\int d t\, \phi_n(t)\phi^\ast_m(t)=\int d t \psi_n(t)\psi^\ast_m(t)=\delta_{nm}$ and $\lambda_n$ is the corresponding decomposition coefficient whose value is ordered such that $1\ge \lambda_0\ge\lambda_1\ge\ldots\ge0$. In terms of these modes, the input-output relation in Eq.~(\ref{eq2}) becomes
\begin{equation}
\label{eq4}
    \hat{b}_n=\lambda_n \hat{a}_n+\hat{\xi}_n, ~~\mathrm{for}~ n=0,1,2,...,
\end{equation}
where $\hat{b}_n=\int d t \,\hat{b}(t)\phi_n^\ast(t)$ and $\hat{a}_n=\int d t \,\hat{a}(t)\psi_n^\ast(t)$ are the annihilation operators for the SF and the signal waves, respectively, in the $n$-th pair of normal modes. $\hat{\xi}_n=\int d t
\,\hat{\xi}(t) \phi_n^\ast(t)$ is a noise operator in the same output mode satisfying $[\hat{\xi}_n,\hat{\xi}^\dag_n]=1-\lambda_n^2$.

From Eq.~(\ref{eq4}), it is clear that the waveguide functions as a multi-port ``beamsplitter'' that converts an input signal in mode $\psi_n(t)$ into an output mode $\phi_n(t)$ in the SF band with probability $\lambda_n^2$. When $\lambda_0\sim 1$ and $\lambda_{n>0}\ll 1$, only a single mode, the fundamental mode, is converted with high efficiency while all the other modes remain unaffected \cite{QPG_SilberhornOE11}. Therefore, by passing the input signal through a sequence of QFC stages, each driven by an appropriately-shaped pump pulse to provide high efficiency up-conversion to a single signal mode that is orthogonal to the rest, MRPC of photonic signals in multiple temporal modes can be implemented utilizing a serial or an optical-loop configuration as shown in Figs.~\ref{fig1}(a) and \ref{fig1}(b), respectively.

\begin{figure}
   \centering \epsfig{figure=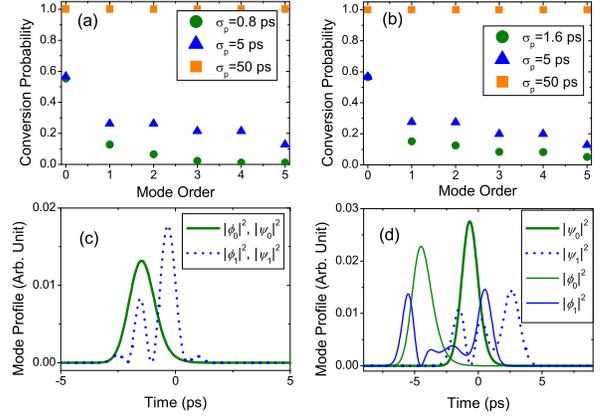,width=7.8cm}
   \caption{Mode structures of QFC. (a) and (b) show the conversion efficiencies of the first few normal modes for $\mu=-\nu=1$ ps/cm and $\mu=3\nu=3$ ps/cm, respectively, driven by Gaussian-shaped pump pulses of different temporal widths. (c) shows the profiles of the major modes in (a) with $\sigma_p=0.8$ ps, whereas (d) shows those in (b) with $\sigma_p=1.6$ ps.}  \label{fig2}
\end{figure}

In practice, single-mode QFC can be achieved by tailoring the dispersion properties of the waveguide. For example, by engineering the group velocity of the pump wave to match that of either the signal or the SF wave while making it very different from the other \cite{TailoringFWM10}, single-mode QFC can be approximately realized \cite{QPG_SilberhornOE11}. This approach, however, is applicable to only a few optical materials and requires complicated experimental procedures for tailoring the waveguide's geometry.

We propose to achieve single-mode QFC by modulating the pump pulses instead, thus avoiding the need for dispersion tailoring of the waveguides while allowing real-time management of the QFC modes. As we have previously demonstrated \cite{HuaAltKum10,PhysRevA.84.033844,PatalPRA12}, entangled photons can be generated in single modes in any nonlinear medium as long as $BT<1$, where $B$ is the phase-matching bandwidth and $T$ is the temporal width of the pump pulses. Intuitively, this is because with $BT<1$, it is impossible, even in principle, to determine the exact temporal location of the generated photons within $T$, so that they must subtend a single coherent mode on the detection apparatus. Applying the same principle to the QFC process, one should expect similar single-mode operation using pump pulses of temporal length comparable to or shorter than the reciprocal of the waveguides's phase-matching bandwidth. This argument should hold for any nonlinear medium with arbitrary dispersion property.

As an example, in Fig.~\ref{fig2} we show the QFC mode structures for a 1-cm-long waveguide pumped by Gaussian pulses of the form $f(t)=e^{-t^2/2\sigma_p^2}$. We choose $\eta=0.5 \pi $/cm so as to achieve complete frequency conversion at the peak of the pump pulses under the condition of perfect phase matching. Note that depending on the application, this may not be the optimum choice. In Fig.~\ref{fig2}(a) we plot the conversion probabilities for the first few normal modes for $\mu=1$ ps/cm and $\nu=-1 $ ps/cm. As shown, when the pump pulses are long, multiple modes are simultaneously converted. However, when they become shorter, single-mode operation of the QFC process can be approximately achieved, where only the fundamental input signal mode $\psi_0(t)$ is efficiently frequency converted to the fundamental mode of the SF wave $\phi_0(t)$. For example, when $\sigma_p=0.8$ ps the conversion efficiency is close to 0.6 for the fundamental mode, while it is only around 0.1 for the 1st-order mode, and nearly zero for all the other modes. The temporal profiles of the major modes are plotted in Fig.~\ref{fig2}(c), where for the chosen parameters the paired input-output modes turn out to have identical profiles. We emphasize that such predominately single-mode operation can be realized in any arbitrarily dispersive medium, provided appropriately profiled pump pulses are used. For example, in Figs.~\ref{fig2}(b) and \ref{fig2}(d) we plot the conversion efficiencies and the profiles of the major normal modes, respectively, for $\mu=3$ ps/cm and $\nu=1$ ps/cm, where similar single-mode behavior is seen. Note that all the above results are obtained with temporally Gaussian-shaped pump pulses. By using non-Gaussian chirped pulses, the conversion efficiency for the fundamental mode can be further increased while making it smaller for all the other modes.

\begin{figure}
 \epsfig{figure=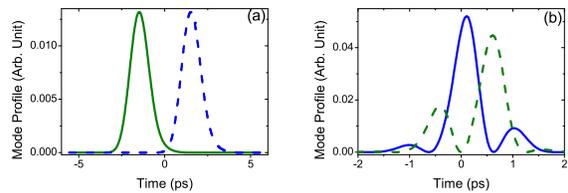,width=7.5cm}
      \caption{(a) Profiles of the fundamental input signal modes for $f(t)=e^{-t^2/2\sigma_p^2}$ (solid) and $f(t)=e^{-(t-3)^2/2\sigma_p^2}$ (dashed) with $\sigma_p=0.8$ ps, $\eta=0.5\pi$/cm, and $\mu=-\nu=1$ ps/cm. (b) Similar profiles for $f(t)=e^{-t^2/2\sigma_p^2} \cos(2t/\sigma_p)$ (solid) and $e^{-t^2/2\sigma_p^2} \sin(2t/\sigma_p)$ (dashed) with $\sigma_p=0.8$~ps, $\eta=0.5\pi$/cm, $\mu=0$, and $\nu=3$~ps. \label{fig3}}
\end{figure}
In order to obtain orthogonal fundamental QFC modes for implementing MRPC, one can simply use temporally delayed pump pulses. For example, two nearly-orthogonal fundamental modes with 2\% overlap are obtained for two identical pump pulses with a relative delay of 3 ps between them, as shown in Fig.~\ref{fig3}(a). Modes of different temporal shapes can be obtained by modulating the amplitude and/or the phase of the pump pulses. For example, Fig.~\ref{fig3}(b) shows two orthogonal fundamental modes with $3\%$ overlap that are obtained for pump pulses of the form $ f(t)=e^{-t^2/2\sigma_p^2} \cos(2t/\sigma_p)$ and $f(t)=1.3 e^{-t^2/2\sigma_p^2} \sin(2t/\sigma_p)$, respectively.

This research was supported in part by the DARPA ZOE program (Grant No. W31P4Q-09-1-0014).

\end{document}